\documentclass[a4paper,10pt]{article}

\usepackage{graphicx,amsfonts}

\begin{document}

% Title, authors and addresses

% use the thanksref command within \title, \author or \address for footnotes;
% use the corauthref command within \author for corresponding author footnotes;
% use the ead command for the email address,
% and the form \ead[url] for the home page:
% \title{Title\thanksref{label1}}
% \thanks[label1]{}
% \author{Name\corauthref{cor1}\thanksref{label2}}
% \ead{email address}
% \ead[url]{home page}
% \thanks[label2]{}
% \corauth[cor1]{}
% \address{Address\thanksref{label3}}
% \thanks[label3]{}

\title{Mixing and approach to equilibrium in the standard
map}

% use optional labels to link authors explicitly to addresses:
% \author[label1,label2]{}
% \address[label1]{}
% \address[label2]{}

\author{F. Baldovin\thanks{E-mail address: baldovin@cbpf.br} \\
   \it{Centro Brasileiro de Pesquisas F\'{\i}sicas,}\\
   \it{Rua Xavier Sigaud 150,
   22290-180 Rio de Janeiro -- RJ, Brazil} }

\date{September 19, 2001}

\maketitle

\begin{abstract}
For a paradigmatic case, the standard map, we discuss how the statistical
description of the approach to equilibrium is related to the sensitivity
to the initial conditions
of the system. Using a numerical analysis we
present an anomalous scenario that may
give some insight on the foundations of the Tsallis' statistical
mechanics.
\end{abstract}



% main text
\section{Introduction}
%\label{}

\begin{figure}
\label{fig_1}
\begin{center}
\includegraphics[width=12cm,angle=0]{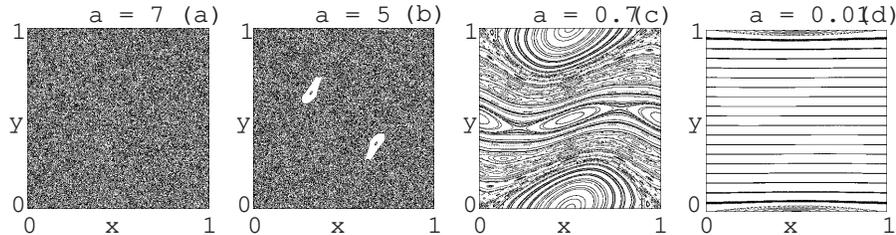}
\end{center}
\caption{\small Phase portrait of the standard map for
typical values of $a$. $N=20\times20$ orbits (black dots)
were started
with a uniform ditribution and traced for $0\leq t\leq 200$.}
\end{figure}

Since long time it is a widely diffused opinion between physicists that the
basis of statistical mechanics lies on dynamics.
Particularly, Krylov \cite{krylov} has pointed out
that the mixing properties of
a dynamical system are responsible for its statistical behaviour (herein
we use this word as a synonym of sensitivity to initial
conditions: $\xi(t)\equiv\lim_{\Delta x(0)\to0}\Delta x(t)/\Delta x(0)$, for
a one-dimensional illustration).
Along this line, intensive work has been done, especially in situations were
anomalous effects may arise (see, for example,
\cite{baldovin} and references therein).
In \cite{baldovin} it was studied numerically
the approach to equilibrium of an Hamiltonian system, the
standard map, also
referred to as the kicked-rotator model:
\begin{eqnarray}
\label{std}
x_{t+1}&=&y_t+\frac{a}{2\pi}\sin(2\pi x_t)+x_t~~~\textup{(mod
1)},\nonumber\\ \\
y_{t+1}&=&y_t+\frac{a}{2\pi}\sin(2\pi x_t)~~~~~~~~~\textup{(mod 1)}\nonumber,
\end{eqnarray}
where $a\in\mathbb{R}$.
This map presents a simplectic structure that
corresponds to an integrable system when $a=0$, while,
for large-enough values of $a$, it is strongly chaotic
(in Fig. \ref{fig_1} we display the phase portrait of the map (\ref{std})
for typical values of $a$).
It was shown that for small values of the parameter $a$
(where the border between the regular and the chaotic region
becomes significant), the approach to equilibrium of the system
displays an anomalous behaviour of the usual, Boltzmann-Gibbs (BG),
statistical entropy; anomaly that may open the door for a
dynamical foundation of the Tsallis' thermodynamics.
In this paper we review the numerical analysis \cite{baldovin},
presenting some new results and making some further speculations.

\section{Statistical description of the approach to equilibrium}
\begin{figure}
\label{fig_2}
\begin{center}
\includegraphics[width=12cm,angle=0]{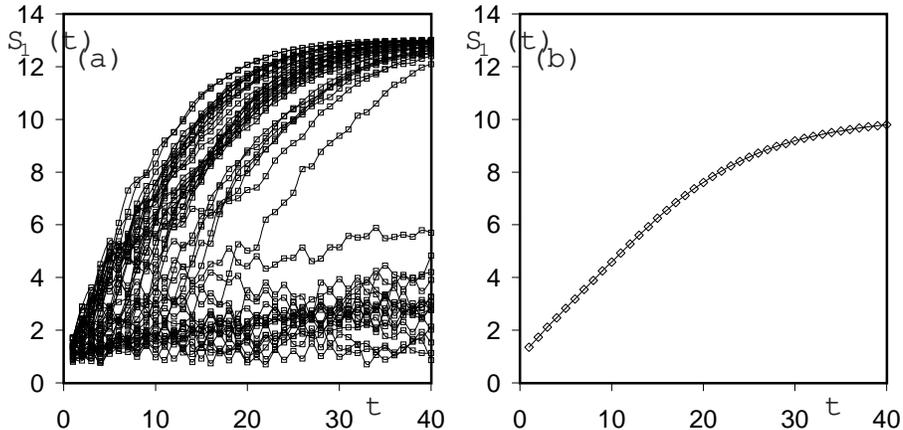}
\end{center}
\caption{\small $S_q(t)$ for $a=2$ and $q=1$, with $N=W=1000\times 1000$
and initial data chosen randomly all over the whole unit square (see text).
(a) $50$ different histories. (b) average over $5000$ histories.}
\end{figure}

Assuming a Gibbsian point of view, we can study the system (\ref{std})
(in its $\Gamma$ space) proceeding, for example, as follows.
We start introducing a coarse-graining
partition of the phase space by dividing it in $W$ cells of equal
size, and we set many copies of the system ($N$ points) in a
far-from-equilibrium situation putting all the $N$ points
inside a single cell.
The occupation number $N_i$ of
each cell $i$ ($\sum_{i=1}^W N_i=N$) provides a probability distribution
$p_i={N_i}/{N}$, hence the definition of an entropy value:
\begin{equation}
\label{ST}
S_q=\frac{1-\sum_{i=1}^W p_{i}^{q} }{q-1}~~~(q\in\mathbb{R}).
\end{equation}
We remind
that the entropic form (\ref{ST})
reduces to the BG entropy
$S_1=-\sum_{i=1}^Wp_i\ln p_i$
in the limit $q\rightarrow 1$
(for a recent review on Tsallis' statistics, see
\cite{tsallis}).
Using then
the dynamic equations (\ref{std}), at each step the points spread in the
phase space causing the entropy value to change.
Finally, to extract a {\em global} quantity on the mapping phase
space, we
repeat the calculation setting the cell that contains the initial points in
different positions chosen randomly all over the {\em whole} unit square, and we
take an average over all the different histories thus obtained.
The key point is to perform many
different histories so that the average stabilizes on a
definite curve.
The result of this
analysis for fixed $a$ is then a single curve of the entropy versus time,
for each entropic
form $S_q$ (in Fig. \ref{fig_2} we show a typical case).

\section{Power-law mixing as a foundation of Tsallis' generalized statistical
mechanics}
We know from the theory of chaos that when a map displays a strong
chaotic behavior, the mixing is exponential. When we attempt an
analysis like the one described in the previous
section, there is just one value of $q$ for which the entropy displays a
linear stage before saturation: $q=1$;
moreover (see \cite{latora1}), the slope of this linear stage is equal to the
Kolmogorov-Sinai entropy rate (i.e., the positive Lyapunov exponent).
The resultant curve has the features of
Fig. \ref{fig_2} (b).
If in the same time interval where $S_1(t)$ grows linearly
we use $q<1$ ($q>1$)  the curve bends
upward (downward) before saturation.
This situation recalls the definition of the Hausdorff's
dimension of a space:
taking {\em first} the limit $W\to\infty$ and {\em then} the limit $t\to\infty$
there is only one value of $q$ for which the entropy production is
different from $0$ and from $+\infty$.

\begin{figure}
\label{fig_3}
\begin{center}
\includegraphics[width=12cm,angle=0]{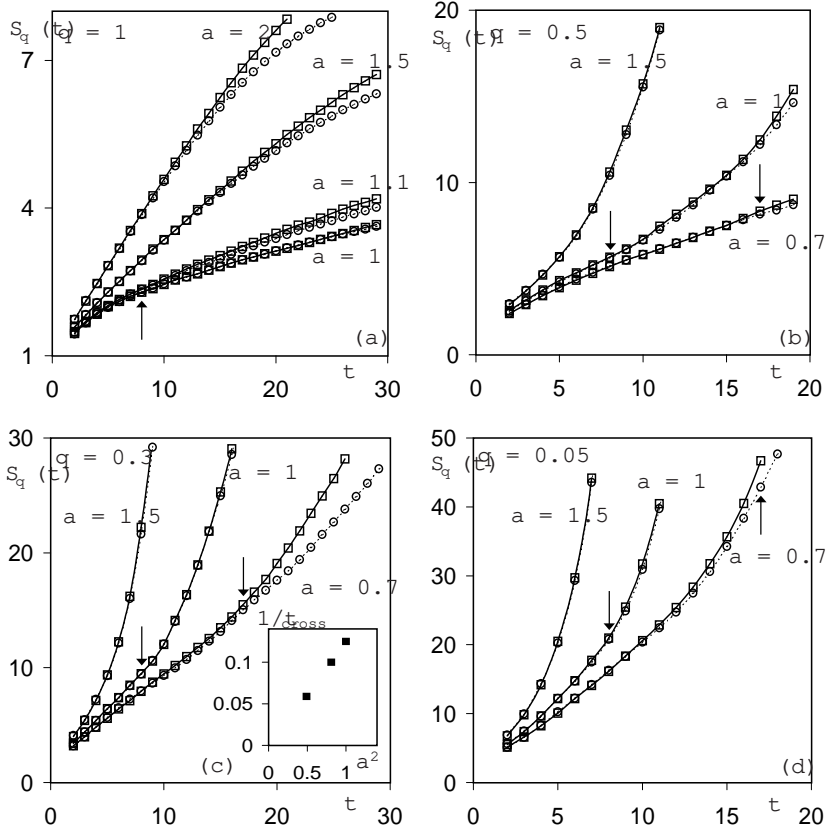}
\end{center}
\caption{\small $S_q(t)$ for typical values of $a$
($5000$ to $16000$ histories were averaged);
squares correspond to $N=W=2236\times 2236$;
circles correspond to $N=W=1000\times 1000$.
The lines are guides to the eye.
(a) $q=1$,  (b) $q=0.5$, (c) $q=0.3$, (d) $q=0.05$
The arrows indicate the crossover time $t_{cross}$ for the corresponding values
of $a$, as represented in the inset of (c).}
\end{figure}

On the other side, if the phase space has characteristics like those in
Fig. \ref{fig_1}(c), the islands-around-islands structure at the
border between the strongly chaotic and the regular regions (see, for
example, \cite{zaslavsky}) slows down the mixing, making it to become a
power-law mixing instead of an exponential one.
In this case,
for time not too large, the linear growth with time occurs
for the entropy $S_{q^*}$ with $q^*<1$ ($q^*\simeq 0.3$ for the standard map),
and not for $S_1$ (see Fig. \ref{fig_3}(a) and \ref{fig_3}(c)).
Waiting enough time, after $t=t_{cross}$,
a crossover to the exponential mixing would occur, due to the rapidity of
the exponential growth with respect to the power-law one.
We have then two different statistical regimes for $0<|a|<1$:
first ($t<<t_{cross}$),
one for which the mixing properties are well described by a Tsallis'
entropy $S_{q^*}$ with $q^*<1$; then ($t>>t_{cross}$),
one for which the mixing properities
are well described by the usual BG entropy.
%If our previous definition is correct in some sense,
%in this situation we should have
%two different statistical regimes. First, one characterized by a
%Tsallis' entropy $S_{q^*}$, with $q^*<1$; then, one described by
%the usual BG entropy.
The interesting point is that for some particular macroscopic conditions
(here represented by the value of the parameter $a$),
the complexification of the phase space
(especially when dimensionality increases)
may increase $t_{cross}$, keeping the system, for at
least some interesting physical observations, in a meta-stable Tsallis'
phase.
This scenario is consistent with some anomalies observed in long range
many-body Hamiltonians \cite{latora} that were presented at this congress.

In Fig. \ref{fig_3} we present, for typical values of $q$,
how this anomalous stage in the entropy growth
makes its appeareance, when one decreases $|a|$ under certain value.
The crossover time $t_{cross}$ increases when $a\to 0$ (see
inset of Fig. \ref{fig_3}(c)).

\section*{Acknowledgments}
C. Tsallis, B. Schulze and E. Pinheiro Borges
are thanked for fruitful discussions.
I have benefitted from partial support by CAPES, FAPERJ, PRONEX and
CNPq (Brazilian agencies). I also acknowledge the general support
of the organizers of NEXT 2001 meeting and the warm hospitality
of Sardegna.


\begin{thebibliography}{00}

% \bibitem{label}
% Text of bibliographic item

% notes:
% \bibitem{label} \note

% subbibitems:
% \begin{subbibitems}{label}
% \bibitem{label1}
% \bibitem{label2}
% If there is a note, it should come last:
% \bibitem{label3} \note
% \end{subbibitems}

\bibitem{krylov} N. Krylov, Nature {\bf 153}, 709 (1994).

\bibitem{baldovin} F. Baldovin, C. Tsallis and B. Schulze, cond-mat/0108501.

\bibitem{tsallis}
C. Tsallis, in {\em Nonextensive Statistical Mechanics and Its Applications},
eds. S. Abe and Y. Okamoto,
Lecture Notes in Physics {\bf 560}, 3 (Springer, Berlin, 2001).

\bibitem{latora1}
V. Latora, and M. Baranger, Phys.Rev. Lett. {\bf 81}, 520
(1999).

\bibitem{zaslavsky}
G.M. Zaslavsky, and B.A. Niyazov, Phys. Rep. {\bf 283}, 73 (1997).

\bibitem{latora}
V. Latora, A. Rapisarda and C. Tsallis, Phys. Rev. E (2001), in press
(cond-mat/0103540).

\end{thebibliography}
\end{document}